\documentclass[12pt, letterpaper]{article}

\usepackage{ifthen}
\newcommand{\Bset}[1]{\setboolean{#1}{true}}
\newcommand{\Bunset}[1]{\setboolean{#1}{false}}

 \usepackage{amssymb}
 \usepackage[dvips]{graphicx}
 \usepackage{amsmath}  
 \usepackage{amsthm}  
 \usepackage{latexsym}
 \usepackage{amsfonts}
 \usepackage{psfrag}   
 \usepackage[usenames,dvipsnames]{color}
 \usepackage{url}   

\newcommand{\Bif}[3]{\ifthenelse{\boolean{#1}}{#2}{#3}}

\newtheoremstyle{remark}{}{}{}{}{\itshape}{:}{ }{\thmname{#1}\thmnumber{ #2}}
\newtheoremstyle{remslide}{}{}{}{}{\itshape}{:}{ }{\thmname{#1}}

   {\theoremstyle{remark}\newtheorem{remark}{Remark}}

\newcounter{tmp_id_cnt}
\newcommand{\nospell}[1]{#1}  

\newcommand{\mydef}[2]{\def#1{#2}}

\newcommand{\newident}[3][*]{\ifthenelse{\equal{*}{#1}}
 {\newcommand{#2}[1][*]
  {\ifthenelse{\equal{*}{##1}}
   {\nospell{\mbox{\Ensuremath{{\mathit{#3}}}}}}
   {\ifthenelse{\equal{b}{##1}}
     {\nospell{\mbox{\Ensuremath{{\mathbf{#3}}}}}}
     {#3}}}}
 {\mydef{#2}{#3}}}

\newcommand{\newmat}[3][*]{\ifthenelse{\equal{*}{#1}}
 {\newcommand{#2}[1][*]
  {\ifthenelse{\equal{b}{##1}}
   {\nospell{\mbox{\Ensuremath{\mathbf{#3}}}}}
   {\ifthenelse{
    \( \equal{*}{##1} \and \not \boolean{in_math_mode} \)
    \or \( \not \equal{*}{##1} \and \boolean{in_math_mode} \)}
    {\nospell{\mbox{\Ensuremath{#3}}}}
    {#3}}}}
 {\mydef{#2}{#3}}}

\newcommand{\newmatop}[2]{\mydef{#1}{\operatorname{#2}}}

\newcommand{\MyMakeTheoMacros}[3]{
 \newcommand{#2}[2][]{\ifthenelse{\equal{}{##1}}
  {\begin{#1} ##2 \end{#1}}
  {\begin{#1} ##2 \label{##1}\end{#1}}}
 \newcommand{#3}[3][]{\ifthenelse{\equal{}{##1}}
  {\begin{#1}{\bf\e{##2}} ##3 \end{#1}}
  {\begin{#1}{\bf\e{##2}} ##3 \label{##1}\end{#1}}}
}

\newcommand{\MyMakeRefMacros}[3]{\newcommand{#1}[2][]
 {\ifthenelse{\equal{}{##1}}{#2~\ref{##2}}{#3~\ref{##1} and~\ref{##2}}}}

\newcommand{\MyMakeEqRefMacros}[3]{\newcommand{#1}[2][]
 {\ifthenelse{\equal{}{##1}}{#2~\eqref{##2}}{#3~\eqref{##1} and~\eqref{##2}}}}

\newcommand{\abstr}[1]{
	\begin{abstract}
		#1
	\end{abstract}}

 \newcommand{\bibentry}[8]{

  \bibitem[\nospell{#8}]{#1} {\textup #3}.
  \newblock \textrm{#4.} \newblock {\em #5, #6}, #7.
 
}

 \newcommand{\inputbib}{

\bibentryPerCom{H. Buhrman}{B}

\bibentry{BBT04_Qua}{Brassard, Broadbent and Tapp}{G. Brassard, A. Broadbent and A. Tapp}{Quantum pseudo-telepathy}{Foundations of Physics, Vol. 35, No. 11}{}{2004}{BBT04}

\bibentry{BJK04_Exp}{Bar-Yossef, Jayram and Kerenidis}{Z. Bar-Yossef, T. S. Jayram and I. Kerenidis}{Exponential separation of quantum and classical one-way communication complexity}{Proceedings of 36th Symposium on Theory of Computing}{pp. 128-137}{2004}{BJK04}

\bibentry{CHTW04}{Cleve, Hoyer, Toner and Watrous}{R. Cleve, P. Hoyer, B. Toner and J. Watrous}{Consequences and limits of nonlocal strategies}{Proceedings of the 19th IEEE Conference on Computational Complexity}{pp. 236-249}{2004}{CHTW04}

\bibentry{GKW04_Qua_com}{Gavinsky, Kempe and de Wolf}{D. Gavinsky, J. Kempe and R. de Wolf}{Quantum communication cannot simulate a public coin}{submitted}{arxiv.org/abs/quant-ph/0411051}{2004}{GKW04}

\bibentry{N91_Pri}{Newman}{I. Newman}{Private vs. common random bits in communication complexity}{Information Processing Letters 39(2)}{pp. 67-71}{1991}{N91}

\bibentry{N99_Op_Lo}{Nayak}{A. Nayak}{Optimal Lower Bounds for Quantum Automata and Random Access Codes}{Proceedings of the 40th Annual Symposium on Foundations of Computer Science}{pp. 369-377}{1999}{N99}

\bibentry{Y83_Lo}{Yao}{A. C-C. Yao}{Lower bounds by probabilistic arguments}{Proceedings of the 24th Annual Symposium on Foundations of Computer Science}{pp. 420-428}{1983}{Y83}

\bibentry{Y03_On_the}{Yao}{A. C-C. Yao}{On the power of quantum fingerprinting}{Proceedings of the 35th Symposium on Theory of Computing}{pp. 77-81}{2003}{Y03}

 }

 \newcommand{\bib}[1][]{

  \begin{thebibliography}{ABCDE00}
    \frenchspacing
    \inputbib
    #1
  \end{thebibliography}
 
}

 \newcommand{\citePerCom}{\cite}
 \newcommand{\bibentryPerCom}[2]{
  \bibitem[\nospell{#2}]{#1} {\textup #1} - \textit{Personal communication}.}

\MyMakeTheoMacros{fact}{\fct}{\nfct}

\MyMakeRefMacros{\fctref}{Fact}{Facts}

 \MyMakeTheoMacros{lemma}{\lem}{\nlem}

\newcommand{\duplem}[2]{\def\my_tmp_id{my_tmp_id_\arabic{tmp_id_cnt}}
 \newtheorem*{\my_tmp_id}{Lemma~\ref{#1}}
 \begin{\my_tmp_id} #2 \end{\my_tmp_id}\stepcounter{tmp_id_cnt}}

\MyMakeRefMacros{\lemref}{Lemma}{Lemmas}

\newcommand{\fakelemref}[1]{Lemma~{#1}}

\MyMakeTheoMacros{corollary}{\crl}{\ncrl}

\MyMakeRefMacros{\crlref}{Corollary}{Corollaries}

\newcommand{\rem}[1]{\begin{remark} #1 \end{remark}}

\MyMakeTheoMacros{proposition}{\prp}{\nprp}

\MyMakeRefMacros{\prpref}{Proposition}{Propositions}

\MyMakeTheoMacros{property}{\prpr}{\nprpr}

\MyMakeRefMacros{\prprref}{Property}{Properties}

 \MyMakeTheoMacros{claim}{\clm}{\nclm}

\MyMakeRefMacros{\clmref}{Claim}{Claims}

\MyMakeTheoMacros{theorem}{\theo}{\ntheo}

\MyMakeRefMacros{\theoref}{Theorem}{Theorems}

\MyMakeTheoMacros{definition}{\defi}{\ndefi}

\MyMakeRefMacros{\defiref}{Definition}{Definitions}

\newcommand{\prf}[2][]{\ifthenelse{\equal{}{#1}}
 {\begin{proof}\renewcommand{\qedsymbol}{$\blacksquare$} #2 \end{proof}}
 {\begin{proof}[Proof of #1]
  \renewcommand{\qedsymbol}{$\blacksquare_{\mbox{\it{\scriptsize{#1}}}}$}
  #2 \end{proof}}
}

\newcommand{\prfsk}[2][]{\ifthenelse{\equal{}{#1}}
 {\begin{proof}[Proof sketch]
  \renewcommand{\qedsymbol}{$\blacksquare$} #2 \end{proof}}
 {\begin{proof}[Proof sketch of #1]
  \renewcommand{\qedsymbol}{$\blacksquare_{\mbox{\it{\scriptsize{#1}}}}$}
  #2 \end{proof}}
}

\newcommand{\sect}[2][]{\ifthenelse{\equal{}{#1}}
 {\section{#2}}
 {\section{#2}\label{#1}}}

\newcommand{\ssect}[2][]{\ifthenelse{\equal{}{#1}}
 {\subsection{#2}}
 {\subsection{#2}\label{#1}}}

\newcommand{\sssect}[2][]{\ifthenelse{\equal{}{#1}}
 {\subsubsection{#2}}
 {\subsubsection{#2}\label{#1}}}

\MyMakeRefMacros{\sref}{Section}{Sections}

\MyMakeRefMacros{\ssref}{Subsection}{Subsections}

\MyMakeRefMacros{\sssref}{Subsection}{Subsections}

\newboolean{in_math_mode}
\setboolean{in_math_mode}{false}

\newcommand{\IfMathMode}{\Bif{in_math_mode}}

\newcommand{\MathModeOn}{\Bset{in_math_mode}}

\newcommand{\MathModeOff}{\Bunset{in_math_mode}}

\newcommand{\Ensuremath}[1]{\IfMathMode  
 {\ensuremath{#1}}
 {\MathModeOn\ensuremath{#1}\MathModeOff}}

\newcommand{\fbr}[1]{\IfMathMode  
 {$#1$}                             
 {\MathModeOn$#1$\MathModeOff}}

\newcommand{\fla}[2][*]{\ifthenelse{\equal{}{#1}}
 {\fbr{#2}}
 {\mbox{\fbr{#2}}}{}}

\newcommand{\bfla}[2][]{\mbox{\fbr{\mathbf{#2}}}}

\newcommand{\mat}[2][]{\ifthenelse{\equal{}{#1}}  
 {\begin{displaymath} \MathModeOn
  #2
  \MathModeOff \end{displaymath}    
 }{
  \begin{equation} \MathModeOn \label{#1}
  #2
  \MathModeOff \end{equation}
 }
}

\newcommand{\matal}[2][]{\mat[#1]{\begin{split} #2 \end{split}}}

\MyMakeEqRefMacros{\equref}{Equation}{Equations}

\MyMakeEqRefMacros{\expref}{Expression}{Expressions}

\MyMakeEqRefMacros{\inequref}{Inequality}{Inequalities}

\MyMakeRefMacros{\figref}{Figure}{Figures}

\newcommand{\poly}{\mathop{\mbox{poly}}}

\newcommand{\Exp}[2][]{\mathop{\mathbf{E}}_{#1}\left[{#2}\right]}

\newcommand{\PR}[2][]{\mathop{\Pr_{#1}\left[{#2}\right]}}

\newcommand{\pl}[1][]{\nospell{\ifthenelse{\equal{}{#1}}
 {\mbox{-s}}
 {\fla{#1}\mbox{-s}}}}

\newmat[]{\llp}{\left(}
\newmat[]{\rrp}{\right)}

\newmat[]{\lla}{\left\langle}
\newmat[]{\rra}{\right\rangle}

\newmat[]{\dt}{\cdot}
\newmat[]{\tm}{\cdot}
\newmat[]{\xor}{\oplus}
\newmat[]{\sbseq}{\subseteq}
\newmat[]{\nsbse}{\nsubseteq}
\newmat[]{\sbs}{\subset}
\newmat[]{\deq}{\stackrel{\textrm{def}}{=}}

\newmat{\NN}{\mathbb{N}}

\newcommand{\quo}[1]{\begin{quote} #1 \end{quote}}

\newcommand{\itemi}[1]{\begin{itemize} #1 \end{itemize}}

\newcommand{\wlg}{w.l.g.\ }	

\newcommand{\st}{such that\ }		

\newcommand{\set}[2][]{\ifthenelse{\equal{}{#1}}
 {\Ensuremath{\left\{#2\right\}}}
 {\Ensuremath{\left\{#2\left|~#1\right.\right\}}}}

\newcommand{\ket}[1]{\Ensuremath{\left|#1\rra}}
\newmat{\kI}{\ket{1}}
\newcommand{\bra}[1]{\Ensuremath{\lla #1\right|}}
\newcommand{\bket}[2]{\Ensuremath{\lla #1\big|#2\rra}}

\newmat[]{\01}{\set{0,1}}
\newcommand{\Zn}[1]{\Ensuremath{[#1]}}

\newcommand{\sz}[2][]{\ifthenelse{\equal{}{#1}}
 {\Ensuremath{\left|#2\right|}}
 {\Ensuremath{\left|#2\right|_{#1}}}}

\newcommand{\asO}[1]{\Ensuremath{O\llp #1\rrp}}
\newcommand{\asOm}[1]{\Ensuremath{\Omega\llp #1\rrp}}

\newcommand{\ceil}[1]{\left\lceil #1 \right\rceil}

\newcommand{\fn}{\footnote}

\newcommand{\nin}{\not\in}  

\newcommand{\f}{\fla}

\newcommand{\fb}{\bfla}

\newcommand{\m}{\mat}

\newcommand{\mal}{\matal}

 \newcommand{\e}{\emph}

\newcommand{\E}{\Exp}

\newcommand{\txt}[1]{\textrm{#1}}  

 \usepackage{hyphenat}  

\date{}
\newmat{\aA}{a^{(A)}}
\newmat{\aB}{a^{(B)}}
\newmat{\xA}{x^{(A)}}
\newmat{\xB}{x^{(B)}}
\newmat{\qA}{q^{(A)}}
\newmat{\qB}{q^{(B)}}

\newmatop{\subm}{sub_m}

\newident{\RI}{R^{1}}
\newident{\RIp}{R^{1}_p}
\newident{\RIe}{R^{1}_e}
\newident{\RII}{R^{\parallel}}
\newident{\RIIp}{R^{\parallel}_p}
\newident{\RIIe}{R^{\parallel}_e}
\newident{\QI}{Q^{1}}
\newident{\QIp}{Q^{1}_p}
\newident{\QIe}{Q^{1}_e}
\newident{\QII}{Q^{\parallel}}
\newident{\QIIp}{Q^{\parallel}_p}
\newident{\QIIe}{Q^{\parallel}_e}

\newident{\MHM}{MHM_n}

\title{A Note on Shared Randomness and Shared Entanglement in Communication}

\author{
  {\bf Dmitry Gavinsky} \\
  {\small Department of Computer Science}\\
  {\small University of Calgary}\\
  {\small Calgary, Alberta, Canada, T2N 1N4}\\
}

\begin{document}

\maketitle

\abstr{We consider several models of 1-round classical and quantum
communication, some of these models have not been defined before. 
We ``almost separate'' the models of \e{simultaneous quantum
message passing with shared entanglement} and the model of \e{simultaneous
quantum message passing with shared randomness}.
We define a relation which can be efficiently \e{exactly} solved in the
first model but cannot be solved efficiently, \e{either exactly or in 0-error
setup} in the second model.
In fact, our relation is exactly solvable even in a more restricted 
model of \e{simultaneous classical message passing with 
shared entanglement}.

As our second contribution we strengthen a result by Yao which says that a 
``very short'' protocol from the model of \e{simultaneous} classical 
message passing with shared randomness can be simulated in the model of 
simultaneous quantum message passing: for a boolean function $f$,
\f[]{\QII(f)\in\exp(O(\RIIp[](f)))\tm\log n.}
We show a similar result for protocols from a (stronger) model of \e{1-way}
classical message passing with shared randomness:
\f[]{\QII(f)\in\exp(O(\RIp[](f)))\tm\log n.}
We demonstrate a problem whose efficient solution in \QII\
follows from our result but not from Yao's.}

\sect{Introduction}
In this work we consider several models of 1-round communication, both
classical and quantum (see \sref{sec_not} for definitions).
To the best of our knowledge, some of these models have never been studied
before, nor were they defined.
For example, we consider the situations when quantum communication channels
are used but the players share a classical random string (and no
entanglement), as well as the case when the players share entanglement but
communication channels are classical.

We argue that these ``unusual'' models are interesting to look at.
There are several open questions related to these models which can be
viewed as natural analogues of yet unsolved questions related to the usual
models.
Besides, models like \e{entangled parties with classical communication
channels}
might correspond to some ``realistic'' circumstances, when physical constrains
make it hard to establish quantum communication.
We also note that the setup of entangled players sending simultaneous
classical messages is closely related to the matters considered by Cleve et
al.\ in \cite{CHTW04}.

Our main interest in this paper is the power of shared
entanglement in communication; in particular, we are looking for similarities
and differences between entanglement and the (intuitively similar) resource
of shared randomness in the context of 1-round communication models.
To the best of our knowledge, prior to this work there was no explicit
construction showing that entanglement can be more powerful than shared
randomness in the same model of communication (for that one has to deal with
at least one model which is \e{unusual} in the above sense). 

\ssect{Our Results}
In this work we ``almost separate'' the models of \e{simultaneous quantum
message passing with shared entanglement} and the model of \e{simultaneous
quantum message passing with shared randomness}.
We define a relation $\MHM$ which can be solved \e{exactly} in the first
model\fn
{Actually, we show that \MHM\ is solvable in a more restricted model \RIIe.}
but cannot be solved \e{either exactly or in 0-error setup} in the
second model (where ``solved'' means efficient solution of cost
\f{\poly(\log n)}).
We \e{conjecture} that our relation is hard for the latter model even in the
bounded error case.

Besides, we extend a result by Yao who has shown that a ``very short''
protocol from the model of \e{simultaneous} classical message passing with
shared randomness can be simulated in the model of simultaneous quantum
message passing (with no shared resource): for a boolean function $f$,
\f[]{\QII(f)\in2^{\asO{\RIIp[](f)}}\tm\log n.}
We show a similar result for protocols from a (stronger) model of \e{1-way}
classical message passing with shared randomness:
for a boolean function $f$, \f[]{\QII(f)\in2^{\asO{\RIp[](f)}}\tm\log n.}
We demonstrate a communication problem whose efficient solution in \QII\
follows from our simulation technique but not from Yao's.

As a straightforward corollary, we show that a protocol of constant cost in
the model of \e{1-way quantum} message passing with shared randomness can be
efficiently simulated in the model of simultaneous quantum message passing. 

\sect[sec_not]{Notation}
In this paper we will consider several models of single round communication
between two parties (\e{Alice} and \e{Bob}).
In general, a communication task can be vied as follows: Alice receives
an input string $x$, Bob receives an input string $y$, then some communication
occurs which allows to compute output, based on $x$ and $y$.
The goal is to produce ``good'' output using minimum amount of communication
(measured in either bits or qubits).
Communication task defines which outputs are good for every possible input,
this task can be given in terms of either a function or a relation 
(the latter allows several good outputs for each input).

A \e{communication model} defines what sort of communication can be performed
in order to solve the problem.
We will consider two types of models.
In \e{simultaneous message passing} models Alice and Bob send one message
each to the third party (a \e{referee}), who has to produce an output
based on the received messages (the 
model is called simultaneous since Alice and Bob do not receive any
information from each other and therefore they can produce their messages
simultaneously, or more precisely, asynchronously).
In \e{1-way communication} models Alice sends a message to Bob and he has to
produce an output based on that message and his part of the input.

Sometimes an additional resource is given to Alice and Bob in order to reduce
the communication cost, that can be either a string of shared
random bits or shared pairs of entangled qubits (w.l.g., EPR pairs).
In either case, the amount of the shared resource is not limited by the
model.

We will be interested in the following models of communication:
\defi{Let
\quo
 {\RI, \RIp, \RIe, \RII, \RIIp, \RIIe, \QI, \QIp, \QIe, \QII, \QIIp, \QIIe}
be 1-round communication models defined as follows: \itemi{
 \item \fb{R} corresponds to communication using classical channels and \fb{Q}
corresponds to communication through quantum channels;
 \item the superscript \fb{\parallel} corresponds to simultaneous message
passing and \fb{1} corresponds to 1-way communication;
 \item the subscript \fb{p} means that a ``public'' random string is shared
between two communicating parties (two broadcasting parties in case of
simultaneous message passing), and the subscript \fb{e} means that EPR
pairs are shared between the parties.}}

We will consider communication complexity of both functions and relations.
We require that a protocol produces right answer with \e{constant advantage
over trivial} (where ``trivial'' usually means achievable by a random guess,
e.g., nontrivial accuracy is any constant greater than $1/2$ in the case of
binary functions, while any constant greater than $0$ is nontrivial in the
case of \f0-error protocols).

Given a communication task $P$ (either a
function or a relation), we will write \f{\RI(P),} etc.\ to address the number
of (qu)bits transferred by the ``cheapest'' protocol solving $P$ in the
corresponding model.
We will write \f{P\in\RI} when \f{\RI(P)\in\poly(\log n),} and
similarly for the rest of defined models.

We will use sign \f{\sbs} to denote proper sets inclusion and by \f{\sbseq}
we will denote the regular inclusion.

\sect{Classical Randomness in Communication}
Quantum communication protocol with (classical) random coins (in both \QIp\
and \QIIp) can be naturally viewed as a ``classical mixture'' of quantum
protocols, where the random string is replaced with a predefined binary
sequence.\fn
{In the case of \QIIp\ there exists one obstacle: the referee does not know
the random string shared between two broadcasting parties.
However we will see that \f{\asO{\log n}} bits of shared randomness are
sufficient, so the referee can receive their values from one of the parties,
at the price of additional \f{\asO{\log n}} communicated qubits.}

That is why several classical (in both senses) results readily extend to the
models \QIp\ and \QIIp.
Let us mention two of them, which we will need later.

\nfct[fct_N]{(\cite{N91_Pri})}{Communication protocols in the models with
shared randomness can \wlg be assumed to be using \f{\asO{\log n}} bits of
shared randomness.}
In particular, it follows that \f{\RIp=\RI,} because Alice can send to Bob
\f{\asO{\log n}} random bits she used, thus making them ``public''.

By \e{distributional deterministic} communication complexity of a problem
with respect to input distribution $D$ we mean the minimum cost of
a deterministic protocol solving the problem with constant advantage over
trivial success probability, when the input is distributed according to $D$.
\nfct[fct_Y]{(\cite{Y83_Lo})}{Communication cost of a problem in the models
with shared randomness is lower bounded by the distributional deterministic
complexity of the same problem in the same model, with respect to any input
distribution.}
Actually, \fctref{fct_Y} is one direction (the ``easy'' one) of the Minimax
Theorem.

\sect{1-Way Communication}
In this section we give two simple equivalence results for the 1-way
communication models we defined.

The first equivalence follows from \fctref{fct_N}:
\prp[eq_p]{\QIp=\QI.}
Similarly to the case of classical communication channels, Alice can send to
Bob \f{\asO{\log n}} random bits, thus making them ``public''.

The second equivalence is a consequence of the quantum teleportation
phenomena:
\prp[eq_e]{\RIe=\QIe.}
Using shared EPR pairs, Alice can teleport to Bob through the classical
channel her quantum message.

\sect[SMP]{Simultaneous Message Passing}
Usually the setting of simultaneous message passing is more interesting (and
sometimes harder to analyze) than 1-way communication.
For example, in the classical theory of communication complexity the model of
simultaneous messages is the only one which can be noticeably strengthened
by adding to it shared randomness.

\ssect{A Generalization of Yao's Simulation}
In \cite{Y03_On_the} Yao shows that a protocol of constant complexity in
\RIIp\ can be simulated by a protocol of complexity \f{\asO{\log n}} in
\QII:

\nfct[fct_Y2]{(\cite{Y03_On_the})}{Let $f$ be a boolean function.
Then \m{\QII(f)\in2^{\asO{\RIIp[](f)}}\tm\log n.}}

Let us generalize his result.
We do that in two steps.

\prp[gen_1]{For any boolean function $f$,
\m{\QII(f)\in2^{\asO{\RIp[](f)}}\tm\log n.}}
\rem{Note that in the context of \prpref{gen_1} the communication models
\RIp\ and \RI\ are \e{not equivalent}, since additive factor \f{\log n}
becomes significant (cf.\ with \fctref{fct_N}).}

\prf[\prpref{gen_1}]{Let $s$ be the communication cost of a protocol for
\f{f(x,y)} in \RIp\ which uses \f{r\in\asO{\log n}} public random bits and is
correct with at least constant probability higher than $1/2$.
Let \f{a(x,q)} be the message sent by Alice in this protocol when her
input is $x$ and the random string is $q$.
Let \f{b(y,a,q)} be a boolean predicate (\01-valued) getting value $1$
if Bob accepts given input $y$, random string $q$ and the message from Alice
being $a$.

To ``simulate'' the original protocol in \QII, do the following (for
\f{k\in\NN} and \f{0<\tau<1} to be chosen later):
\itemi{
 \item Alice sends $k$ copies of
\m{\ket\alpha\deq2^{-\frac r2}\dt\sum_q\ket q\ket{a(x,q)}\kI.}
 \item Bob sends $k$ copies of
\m{\ket\beta\deq2^{-\frac{r+s}2}\dt\sum_{q,a}\ket q\ket a\ket{b(y,a,q)}.}
 \item Using the swap test, the referee estimates the value of
\f{\bket{\alpha}{\beta}} and accepts if the approximation is at least $\tau$.}

Observe that
\mal{\bket{\alpha}{\beta}
 &=2^{-(r+\frac s2)}\sum_{q,a}\bket qq\bket{a(x,q)}a\bket1{b(y,a,q)}\\
 &=2^{-\frac s2}\E[q]{I(q)},}
where \f{I(q)} is an indicator of the event \e{given the random
string $q$, Alice sends a message which causes Bob to accept}.
In other words, \f{I(q)} gets value $1$ if and only if the content $q$ of the
shared random string would cause the original protocol to accept.

Therefore, if we set \f{\tau=2^{-(s/2+1)}} and \f{k\in2^{\asO{s}}} high
enough, with high confidence the new protocol would accept \f{(x,y)} if and
only if the original one would do so with probability greater than $1/2$.
The cost of the new protocol is \f{2^{\asO s}\tm\log n,} as required.}

\prp[gen_2]{For any function $f$,
\m{\RIp(f)\in2^{\asO{\QIp[](f)}}.}}

\prfsk[\prpref{gen_2}]{Assume \wlg that the messages from Alice
are pure states (this can be achieved by at most doubling the number of
communicated qubits).
Since we only require constant precision, we can replace the quantum message
from Alice by a classical one of exponential length.}

Based on \prpref[gen_1]{gen_2}, we get the following corollary:
\crl[gen_3]{A function $f$ of constant communication complexity in \QIp\ can
be solved in \QII\ using \f{\asO{\log n}} qubits of communication.}

\sssect{Strength of Our Improvement}
We think that \prpref{gen_2} and \crlref{gen_3} are not very interesting from
the technical point of view.
In particular, \crlref{gen_3} might be established without \prpref{gen_1},
based on the (trivial) statement that \e{for any boolean $f$,
\f{\RIIp(f)\in2^{\asO{\RIp[](f)}}}} and on the original \fctref{fct_Y2}.

On the other hand, \prpref{gen_1} is less trivial.
Note that if we would try to establish a similar result through application of
\fctref{fct_Y2} and simulating an \RIp-protocol in \RIIp, the price we would
pay for such simplification would be exponential loss in tightness (we 
would end up with something like 
\f{\QII(f)\in\exp\llp\exp{\asO{\RIp[](f)}}\rrp\tm\log n}).

To establish the ``usefulness'' of \prpref{gen_1} let us define a
function $f$ such that
\m{\RIp(f)\in\asO{\log(\log n)}}
but 
\m{\RIIp(f)\in\asOm{\log n}.}
In other words, membership of $f$ in \QII\ would follow from \prpref{gen_1},
but not from Yao's original \fctref{fct_Y2}.

We first define an ``auxiliary'' predicate.
\defi[def_sub]{Let \f{x\in\01^{\log\log m}} and \f{y=(y_1,..,y_{(\log m)/2}),}
where all \pl[y_i] are distinct binary strings of length \f{\log\log m.}
Then \f{\subm(x,y)=1} if and only if $x$ is identical to one of \pl[y_i].}

\defi[def_f]{Let \f[]{a=(a_1,...,a_m)} and 
\f[]{b=(b_1,...,b_m),} where each pair \f{(a_i,b_i)} forms a correct input to
$\subm$.
Then \f{f(a,b)=1} if \f[]{\sz{\set[\subm(a_i,b_i)]{i}}\ge m/2} and
\f{f(a,b)=0} if \f[]{\sz{\set[\subm(a_i,b_i)]{i}}=0}, with a promise that
one of the two cases holds.}

As usual, we use $n$ to denote the length of input to $f$.

To see that \f{\RIp(f)\in\asO{\log(\log n)},} consider a protocol where
several random indices in the range \set{1,..,m} are tossed as a public coin,
then Alice sends \pl[a_i] corresponding to those random indices and Bob
accepts if and only if for at least one of the indices \f{\subm(a_i,b_i)} is
satisfied.

To show that \f{\RIIp(f)\in\asOm{\log n}} we use an approach suggested by
\fctref{fct_Y}, we will fix the input distribution and restrict our attention
to deterministic protocols.
Denote by $X$ the set of \pl[(a,b)] which form correct input to $f$ and in
every \f{a=(a_1,...,a_m)} all \pl[a_i] are identical and in every
\f{b=(b_1,...,b_m)} all \pl[b_i] are identical.
Then according to our distribution, the input is with probability $1/2$ a
uniformly chosen positive instance from $X$ and with probability $1/2$ a
uniformly chosen negative instance from $X$.

Under this distribution the actual task of a protocol solving $f$ would be to
solve an instance of \f{\subm(a_0,b_0)} when Alice receives 
\f{a_0} and Bob receives \f{b_0.}
Our input distribution for \f{\subm(a_0,b_0)} corresponds to uniformly choosing
a positive instance with probability $1/2$ and a negative
instance with probability $1/2$.

Let us see that this task requires \f{\asOm{\log n}} communication in the
model of \RII, when the protocol is deterministic (and the error is bounded by
a constant smaller than $1/2$).
W.l.g., let Alice always send $a_0$ to the referee.
Then it is ``easy to see'' that Bob has to send \f{\asOm{\log n}} bits, since
his part of input is a random subset of \set{1,..,m} of size \f{m/2} and his
message to the referee should enable the recipient to decide with constant
accuracy whether a random $a_0$ belongs to that subset.

\rem{Note that we could use a ``padded version'' of $\subm$ as a
communication task whose membership in \QII\ follows from \prpref{gen_1},
but not from \fctref{fct_Y2}.
However, the example of $f$ is probably more interesting, since for its
efficient solution shared randomness is necessary (while for solving $\subm$
shared randomness is not required).}

\rem{There exists another qualitative difference between our \prpref{gen_1}
and Yao's \fctref{fct_Y2}: while the latter easily extends to the case of
\e{relational problems} (\cite{GKW04_Qua_com}), the bound obtained in
\prpref{gen_1} seems to be ``inherently functional'' (even boolean, in some
sense).}

\ssect{Separating \QIIe\ from \QIIp\ and \RIIe\ from \RIIp}
In this section we ``almost show'' that \f{\RIIe\nsbse\QIIp} (and 
therefore \f{\QIIp\sbs\QIIe}).
We demonstrate a relation which can be solved \e{exactly} in \RIIe (and 
therefore in \QIIe), but cannot be efficiently
solved in \QIIp\ in the following setting: the referee may not
make a mistake, however he is allowed to announce \e{don't know} with some
constant probability (less than $1$).
Of course, this requirement (we call it \e{``don't know'' setting}) is less
severe than exact solvability, and therefore our result may be viewed as
exponential separation between the model of quantum communication with shared
entanglement and the model of quantum communication with shared randomness.

On the other hand, we \e{conjecture} that our relation is hard for the
``standard'' \QIIp\ (which we address as \e{bounded-error setting}), but
we could not prove that.
However, there is a known lower bound for the bounded-error setting for 
the model \RI\ (and therefore for \RIIp) given in \cite{BJK04_Exp} for a 
communication problem which is actually a simplified (and easier for 
communication) version of the relation we will define here.
So, we conclude that \f{\RIIp\sbs\RIIe} and even that \f{\RIIe\nsbse\RI} 
in the standard bounded-error setting.\fn
{The last observation is actually not a contribution of this paper
but rather a \e{compilation} of lower and upper bounds from 
\cite{BJK04_Exp} and \citePerCom{H. Buhrman}, correspondingly.}

Generalizing a construction used in \cite{BJK04_Exp}, we
define a family of relations parametrized by 
\set[n\txt{ is even}]{n\in\NN}.
Denote by \f{M_n} the set of all perfect matchings over $n$ elements 
(labeled as \set{1,...,n}).
\defi{Let \f{a\in\01^n} and \f{m\in M_n.}
For any \f{x=(\aA,m)} and \f{y=\aB} \st\ 
\f{\aA\xor\aB=a} (where $\xor$ means bit-wise xor),
\m{\MHM(x,y)=\set[a_i\xor a_j=b,~(i,j)\in m]{(i,j,b)}.}}

In order to show that \MHM\ is efficiently exactly solvable in \RIIe\ we 
adapt a clever protocol suggested by Buhrman \citePerCom{H. Buhrman} 
(see also \cite{BBT04_Qua}).

\sssect[ss_inQIIe]{\MHM\ Is Exactly Solvable in \RIIe}
Consider the following \RIIe-protocol for \MHM.
\itemi{
 \item Before the communication starts, Alice and Bob share \f{\ceil{\log n}}
pairs of entangled qubits: \f{\sum_{i\in\Zn n}\ket i\ket i.}
 \item When Alice receives \f{x=(\aA,m)} she applies the following
transformation to her part of the entangled pairs:
\m{\ket i\to(-1)^{\aA_i}\ket i.}
Similarly, Bob flips the sign of those parts of the entangled 
\f{\sum\ket i\ket i} which correspond to \f{\aB_i=1.}
 \item Alice performs partial projection of her part of entangled qubits 
to the subspaces of dimension $2$ spanned by all pairs of indices in the 
matching $m$ (fulfilled if necessary by ``insignificant'' subspace of 
the indices greater than $n$, less or equal to \f{2^{\ceil{\log n}}}).
After the measurement, the common state of shared entangled pairs becomes
\m{\frac1{\sqrt2}(\ket k\ket k\pm\ket l\ket l),}
where \f{(k,l)\in m.}
Moreover, the amplitudes of $\ket k\ket k$ and $\ket l\ket l$ coincide if 
and only if \f{a_k=a_l.}
 \item Both Alice and Bob perform \f{\ceil{\log n}}-qubit Hadamard 
transform on their parts of entangled qubits and then measure in the 
standard basis.
As a result, Alice obtains $b_1$ and Bob obtains $b_2$, such that
\m[ref_gets]{(b_1\xor b_2)\dt(k\xor l)=a_k\xor a_l,}
where $\xor$ denotes bit-wise xor operation and $\dt$ stands for the inner 
product $mod$ $2$ of two vectors.
Alice sends \f{(k,l,b_1)} and Bob sends $b_2$ to the referee. 
 \item According to \eqref{ref_gets}, the referee computes \f{a_k\xor a_l} 
and outputs the triple \f{(k,l,a_k\xor a_l),} as required.}

The protocol and the analysis are very similar to those in \cite{BJK04_Exp}.
The probability that the referee produces \f{(k,l,b)} is
\m{\sz{\frac1{\sqrt2}(\bra k+(-1)^b\bra l)\ket\varphi}^2=
 \frac1{2n}((-1)^{a_k}+(-1)^{a_l+b})^2,}
which equals \f{2/n} if \f{(k,l,b)\in\MHM(x,y)} and $0$ otherwise.
The protocol is always correct.

The fact that \f{\MHM\in\RIIe} (which is strengthening of the more obvious 
\f{\MHM\in\QIIe}) is due to \citePerCom{H. Buhrman}.

\sssect{\f{\MHM\nin\QIIp} for the ``Don't Know'' Setting}
Let us slightly simplify the task and assume that the matching $m$ in the 
input to \MHM\ always comes from a fixed family $M'_n$ of $n/2$ 
edge-disjoint matchings on $n$ elements (as before, $n$ is even).
Now we use \fctref{fct_Y} to ``get rid'' of the shared randomness.
We choose the input distribution $D_1$ to be uniform: \f{\aA,\aB\in\01^n,} 
\f{m\in M'_n.}

We will prove the following.
\clm[cl_QII]{The communication cost of computing \MHM\ in the model \QII\ in
the ``don't know'' setting is \f{\asOm{n^{1/6}}}, when the input
distribution is $D_1$.}

The claim leads to the following theorem.
\theo[theo_QII]{The communication cost of computing \MHM\ in the model \QIIp\
in the ``don't know'' setting is \f{\asOm{n^{1/6}}}.}

To prove the claim, we generalize a technique from \cite{GKW04_Qua_com}.
\prf[\clmref{cl_QII}]{Consider a protocol $T$ solving the problem in 
the ``don't know'' setting with success probability at least $7/8$, let 
$s$ be its communication cost (note that in the ``don't know'' setting the 
success probability can be amplified using parallel repetition).

We know that \f{|M'_n|=n/2,} denote by $M_1$ the subset of those 
\f{m\in M'_n} satisfying \f{\PR[D_1]{T \txt{ fails}|m}\le1/4,} where the 
conditioning is on the fact that $m$ is the matching given to Alice.
It must hold that
\m{|M_1|\ge n/4.}

We want to claim that for some matching \f{m\in M_1} the referee outputs
every pair from $m$ with rather low probability, unless $s$ is big.

Let us forget for a moment about the cost of communication
between Alice and the referee, and assume that Alice forwards her complete
input to the referee (this setup corresponds to the 1-way communication model).
Then the referee can output a pair \f{(i,j)\in m} if and only if he knows the
value of \f{\aB_i\xor \aB_j.}

Consider the mixed state corresponding to the uniform distribution of
\f{\aB} (as imposed by $D_1$), and let \f{\beta_x} be the density matrix of
the message sent by Bob to the referee when \f{\aB=x.}
Then the whole message sent to the referee is
\m{\beta=\frac1{2^n}\sum_{x}\beta_x.}

Let us denote by \f{p(i,m)} the probability that the referee outputs the pair
\f{(i,j)\in m} when he receives $\beta$ from Bob and the required matching is
\f{m\in M_1.}
Let
\m{\lambda_m\deq\max_ip(i,m)}
be the highest probability for outputting any specific pair when the required
matching is $m$.
Denote \f{\lambda_0=\min_m\lambda_m.}
For each \f{m\in M_1} let $i_m$ be any fixed index satisfying
\f{p(i_m,m)\ge\lambda_0} and $j_m$ be such that \f{(i_m,j_m)\in m.}

Suppose now that Bob sends \f{1/\lambda_0} independent copies of his message,
then the referee is able to learn the value of \f{\aB_{i_m}\xor\aB_{j_m}}
with probability at least $1/2$. 
In other words, by sending \f{s/\lambda_0} qubits Bob lets the referee learn
any \set[m\in M_1]{\aB_{i_m}\xor\aB_{j_m}} with probability at least $1/2$
(since Bob does not know $m$, his message must be good for any \f{m\in M_1}).
Let \f{E=\set[m\in M_1]{(i_m,j_m)}} and $V$ be the set of endpoints of $E$.
The size of $E$ is at least $n/4$ (since we have required that $M_1$ consists
of disjoint matchings) and \f{|V|\ge\sqrt n/2.}
Let \f{V'\sbs V} be a subset of size \f{\le|V|/2,} such that every point in
\f{V\setminus V'} has at least one neighbor in $V'$ (the neighborhood is
defined by $E$).

Let us further strengthen the referee and let him know the values of all
\set[i\in V']{\aB_i}.
Then by choosing one of \f{m\in M_1} and trying to learn the corresponding 
value of\f{\aB_{i_m}\xor\aB_{j_m}} using the original protocol, the referee is 
able to get (with certainty) the
value of any \set[i\in V\setminus V']{\aB_i} with probability at least $1/2$,
using the message of length \f{s/\lambda_0.}
If instead of announcing ``don't know'' the referee would make a random guess
of the value of \f{\aB_i}, his total correctness probability would increase to
$3/4$.
Since \f{|V\setminus V'|\ge|V|/2\ge\sqrt n/4,} by applying Nayak's bound on
the length of a random access codes (\cite{N99_Op_Lo}) we obtain the
following relation: 
\f[]{s/\lambda_0\ge(1-H(3/4))\tm\sqrt n/4,}
or
\m{\lambda_0\le\frac{22s}{\sqrt n}.}
Fix \f{m_0} to be some matching from $M_1$ satisfying
\f{\max_ip(i,m_0)\le\frac{22s}{\sqrt n}.}

Now let us return to the simultaneous message model.
Denote by $D_2$ the following input distribution: Alice receives
\f{(\aA,m_0)} and Bob receives \aB, where $\aA$ and $\aB$ are
uniformly and independently distributed over \f{\01^n.}
As follows from the definition of $M_1$, the protocol $T$ solves \MHM\ over
$D_2$ with success probability at least $3/4$, and every individual pair from
$m_0$ is returned by $T$ with probability not higher than
\m{p_0\deq\frac{22s}{\sqrt n}.}

Fix an indexing for the edges of the matching we've chosen: 
\f[]{m_0=\set[1\le k\le n/2]{e_k}.}
For the rest of the proof we will use the following notation:
\m{\xA_k=\aA_{i_k}\xor\aA_{j_k},~\xB_k=\aB_{i_k}\xor\aB_{j_k},}
where \f{1\le k\le n/2} and \f{e_k=(i_k,j_k).}
In these terms, given an input from the support of the distribution $D_2,$ 
if the protocol is successful it should produce \f{\xA_k\xor\xB_k} 
together with \f{e_k=(i_k,j_k).}
If the input is distributed according to $D_2$ then the protocol succeeds
with probability at least $3/4$ and each $e_k$ is returned with probability
at most $p_0$.

The latter statement can be reformulated in a stronger form: the information
received by the referee \e{does not allow it} to learn with $0$-error any 
of $e_k$ with probability higher than $p_0$ (this is true since from the 
beginning we might restrict attention to the protocols which always try to 
``boost'' one of the edges).
Let \f{\qA_k} denote the success probability of the referee in guessing
\f{\xA_k} in the ``don't know'' setting, when \aA\ is uniformly chosen.
Then according to the Holevo bound,
\m[m1]{\sum_{1\le k\le n/2}\qA_k\le s.}
Similarly define \f{\qB_k}, then
\m[m2]{\sum_{1\le k\le n/2}\qB_k\le s.}

Now we need the following lemma, which is a simple modification of
\fakelemref{1} from \cite{GKW04_Qua_com} (a proof can be found in the
appendix section). 
\lem[lem_K]{Consider two quantum registers,
the first containing either $\alpha_0$ or $\alpha_1$ and the second
containing either $\beta_0$ or $\beta_1$
(each register contains either one of the corresponding mixed states with
probability $1/2$ and the registers are not correlated in any way).
Denote by $a$ the maximum success probability of distinguishing $\alpha_0$
and $\alpha_1$ in the ``don't know'' setting, define $b$ similarly for
$\beta_0$ and $\beta_1$. 
Let $p$ be the maximum success probability for a
measurements with 3 outcomes:
\f{[\alpha_0\otimes \beta_0 \txt{ or } \alpha_1\otimes \beta_1]},
\f{[\alpha_0\otimes \beta_1 \txt{ or } \alpha_1\otimes \beta_0]}
and $[$don't know$]$ (where ``successful'' are the first two outcomes).
Then \f{p\leq 4ab}.}

According to the lemma, the probability of the
protocol to be successful is upper bounded by
\m{\sum_{1\le k\le n/2}4\qA_k\qB_k,}
subject to \eqref{m1}, \eqref{m2}, and 
\f{\qA_k\qB_k\le p_0} for all \f{1\le k\le n/2} (which follows from the 
obvious lower bound \f{ab} on the success probability in 
simultaneous $0$-error guessing of the content of the both registers in 
the condition of \lemref{lem_K}).
The maximum is achieved if we pick a set $K$ of size \f{s/\sqrt{p_0}} and
fix \f{\qA_k=\qB_k=\sqrt{p_0}} for every \f{k\in K.}
This gives the following upper bound on the success probability:
\m{\frac{s}{\sqrt{p_0}}\dt4p_0=4s\sqrt{p_0}\in
 \asO{\frac{s^{3/2}}{n^{1/4}}},}
and its value must be at least $3/4$.
This leads to \f{s\in\asOm{n^{1/6}}.}}

\sect{Open Problems}

\itemi{
 \item We know that \f{\QI=\QIp\sbseq\QIe=\RIe.} 
Are they equal?
 \item We were only able to demonstrate that \f{\MHM\nin\QIIp} for ``don't
know'' setting, while we conjecture that the problem is hard for $\QIIp$ in
the standard (bounded-error) setting as well.
 \item We have shown our separation using a relation.
Can similar results be obtained for a (partial) boolean function?
What about a total function?
}

\subsection*{Acknowledgments}
I am very grateful to Richard Cleve and Ronald de Wolf for helpful 
discussions, to Richard for valuable references and to Ronald for his 
close proofreading of a draft of this work.
I am also grateful to Harry Buhrman for his kind permission to use his 
construction for showing that \f{\MHM\in\RIIe} (improving upon 
\f{\MHM\in\QIIe} in the preliminary version of this paper).

\bib

\newpage
\appendix

\sect{Appendix}

\ssect{Proof of \lemref{lem_K}}
\duplem{lem_K}{Consider two quantum registers,
the first containing either $\alpha_0$ or $\alpha_1$ and the second
containing either $\beta_0$ or $\beta_1$
(each register contains either one of the corresponding mixed states with
probability $1/2$ and the registers are not correlated in any way).
Denote by $a$ the maximum success probability of distinguishing $\alpha_0$
and $\alpha_1$ in the ``don't know'' setting, define $b$ similarly for
$\beta_0$ and $\beta_1$. 
Let $p$ be the maximum success probability for a
measurements with 3 outcomes:
\f{[\alpha_0\otimes \beta_0 \txt{ or } \alpha_1\otimes \beta_1]},
\f{[\alpha_0\otimes \beta_1 \txt{ or } \alpha_1\otimes \beta_0]}
and $[$don't know$]$ (where ``successful'' are the first two outcomes).
Then \f{p\leq 4ab}.}

\prf[\lemref{lem_K}]{In order to prove the lemma it suffice to slightly
modify the proof of similar \fakelemref{1} in \cite{GKW04_Qua_com}. 

Let us define several subspaces of our quantum registers.
Call \f{S_0} the support of $\alpha_0$ (i.e., the span of its non-zero
eigenvectors), and let \f{S_1} be the support of $\alpha_1$.
Similarly define \f{T_0} and \f{T_1} for $\beta_0$ and $\beta_1$.

According to 0-error requirement of our lemma, the positive
semidefinite operator corresponding to the measurement outcome
\f{[\alpha_0\otimes \beta_0 \txt{ or } \alpha_1\otimes \beta_1]} should have
all its eigenvectors lying inside 
\f[]{R\deq(S_1\times T_0)^\bot\cap(S_0\times T_1)^\bot.}
Expanding, we get
\f[]{R=(S_1^\bot\times T_1^\bot)\xor(S_0^\bot\times T_0^\bot)}
(where $\xor$ denotes direct sum of vector spaces),
so that the probability to observe this outcome is at most the probability to
project the measured state to either \f{S_1^\bot\times T_1^\bot} or
\f{S_0^\bot\times T_0^\bot.}

Similarly, the probability to obtain the outcome
\f{[\alpha_0\otimes \beta_1 \txt{ or } \alpha_1\otimes \beta_0]} is at most
the probability to project the state to either 
\f{S_1^\bot\times T_0^\bot} or \f{S_0^\bot\times T_1^\bot.}

Let $p_0$ be the maximum among the probabilities to project the content of
our two registers to \f{S_i^\bot\times T_j^\bot,} where \f{i,j\in\01,} and
w.l.g., let the maximum be obtained for \f{i=j=0.}
Observe that in this case $p_0$ is just the probability to project the first
register to \f{S_0^\bot} times the probability to project the second register
to \f{T_0^\bot}.
The former is equal to the probability to declare with certainty that ``the
first register contains $\alpha_1$'' and the latter is equal to the probability
to declare ``the second register contains $\beta_1$'', with certainty too.
By definition, this is bounded above by \f{ab,}
and therefore \f{p\le4p_0\le4ab.}}

\end{document}